# Associative Array Model of SQL, NoSQL, and NewSQL Databases

Jeremy Kepner[1,2,3], Vijay Gadepally[1,2], Dylan Hutchison[4], Hayden Jananthan[3,5],
Timothy Mattson[6], Siddharth Samsi[1], Albert Reuther[1]

[1]MIT Lincoln Laboratory, [2]MIT Computer Science & AI Laboratory, [3]MIT Mathematics Department, [4]University of Washington Computer Science Department, [5]Vanderbilt University Mathematics Department, [6]Intel Corporation

*Abstract*—The success of SQL, NoSQL, and NewSQL databases is a reflection of their ability to provide significant functionality and performance benefits for specific domains, such as financial transactions, internet search, and data analysis. The BigDAWG polystore seeks to provide a mechanism to allow applications to transparently achieve the benefits of diverse databases while insulating applications from the details of these databases. Associative arrays provide a common approach to the mathematics found in different databases: sets (SQL), graphs (NoSQL), and matrices (NewSQL). This work presents the SQL relational model in terms of associative arrays and identifies the key mathematical properties that are preserved within SQL. These properties include associativity, commutativity, distributivity, identities, annihilators, and inverses. Performance measurements on distributivity and associativity show the impact these properties can have on associative array operations. These results demonstrate that associative arrays could provide a mathematical model for polystores to optimize the exchange of data and execution queries.

## I. INTRODUCTION

Relational or SQL (Structured Query Language) databases [Codd 1970, Stonebraker 1976] such as PostgreSQL, MySQL, and Oracle have been the de facto interface to databases since the 1980s (see Figure 1) and are the bedrock of electronic transactions around the world. More recently, key-value stores (NoSQL databases) such as Google BigTable [Chang 2008], Apache Accumulo [Wall 2015], and MongoDB [Chodorow 2013] have been developed for representing large sparse tables to aid in the analysis of data for Internet search. As a result, the majority of the data on the Internet is now analyzed using key-value stores [DeCandia et al 2007, Lakshman & Malik 2010, George 2011]. In response to similar performance challenges, the relational database community has developed a new class of databases (NewSQL) such as C-Store [Stonebraker 2005], H-Store [Kallman 2008], SciDB [Balazinska 2009], VoltDB [Stonebraker 2013], and Graphulo [Hutchison 2015] to support new analytics capabilities *within* a database. The SQL, NoSQL, and NewSQL concepts have also been blended in hybrid processing systems, such as Apache Pig [Olston 2008], Apache Spark [Zaharia 2010], and HaLoop [Bu 2010]. An effective mathematical model that encompasses the concepts of SQL, NoSQL, and NewSQL would enable their interoperability. Such a mathematical model is the primary goal of this paper.

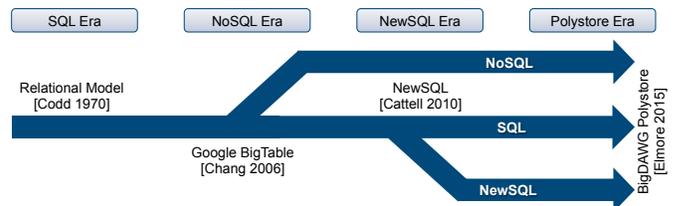

Figure 1. Evolution of SQL, NoSQL, NewSQL, and polystore databases. Each class of database delivered new mathematics, functionality, and performance focused on new application areas.

SQL, NoSQL, and NewSQL databases are designed for specific applications, have distinct data models, and rely on different underlying mathematics (see Figure 2). Because of their differences, each database has unique strengths that are well suited for particular workloads. It is now recognized that special-purpose databases can be 100x faster for a particular application than a general-purpose database [Kepner 2014]. In addition, the availability of high performance data analysis platforms, such as the MIT SuperCloud [Reuther 2013, Prout 2015], allows high performance databases to share the same hardware platform without sacrificing performance.

|  | SQL | NoSQL | NewSQL | Polystore |
|---|---|---|---|---|
| Example | PostgreSQL | Accumulo | SciDB | BigDAWG |
| Application | Transactions | Search | Analysis | All |
| Data Model | Relational Tables | Key-Value Pairs | Sparse Matrices | Associative Arrays |
| Math | Set Theory | Graph Theory | Linear Algebra | Associative Algebra |
| Consistency | ✓ |  |  | ✓ |
| Volume |  | ✓ | ✓ | ✓ |
| Velocity |  | ✓ | ✓ | ✓ |
| Variety |  | ✓ |  | ✓ |
| Analytics |  |  | ✓ | ✓ |
| Usability | ✓ |  |  | ✓ |

Figure 2. Focus areas of SQL, NoSQL, NewSQL, and Polystore databases.

This material is based upon work supported by the National Science Foundation under Grant No. DMS-1312831. Any opinions, findings, and conclusions or recommendations expressed in this material are those of the author(s) and do not necessarily reflect the views of the National Science Foundation.

The recognition of "one size does not fit all" [Stonebraker & Çetintemel 2005] has led to the need for polystore databases, such as BigDAWG [Duggan 2015, Elmore 2015], that can contextualize queries and cast data between multiple databases so that a user can employ the best database for a particular task (see Figure 3). To achieve this goal, polystore databases need to bridge SQL, NoSQL, and NewSQL databases. The Dynamic Distributed Dimensional Data Model (D4M) technology [Kepner 2012] was developed to provide a linear algebraic interface to graphs stored in NoSQL databases [Byun 2012, Kepner 2013]. Subsequently, D4M has been successfully used with both SQL [Wu 2014, Gadepally 2015] and NewSQL [Samsi 2016] databases. The effectiveness of D4M to seamlessly interact with these diverse databases rests on its associative array algebra [Kepner & Chaidez 2013, Kepner & Chaidez 2014, Kepner & Jansen 2016] that provides a mathematics that spans sets, graphs, and matrices. The ability of D4M (and Myria [Halperin 2014]) to bridge multiple databases has laid the foundation for the polystore database concept.

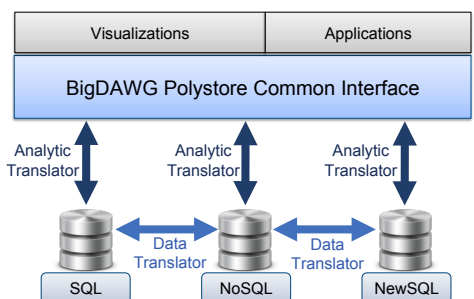

Figure 3. BigDAWG polystore database architecture. Analytic translators contextualize queries to specific databases. Data translators cast data between databases.

Mathematics is one of the most important differences among SQL, NoSQL, and NewSQL databases (see Figure 4). The relational algebra found in SQL databases is based upon selection, union, and intersection of special sets called relations. NoSQL is designed for analyzing sparse relationships among data and relies on graph theory and graph analysis. NewSQL databases use matrices and linear algebra to look for patterns in numeric data.

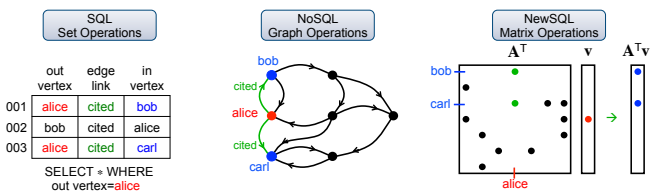

Figure 4. Mathematics of breadth-first search for SQL, NoSQL, and NewSQL databases.

The approach to developing an associative array model of the above databases is as follows. First, the relevant aspects of relations are summarized. Second, the sparse matrix operations that encompass graph algorithms and matrix mathematics are given. Third, the associative array model that describes NoSQL and NewSQL databases is described. Fourth, relations and their corresponding operations are defined in terms of associative arrays. Fifth, the mathematical properties required by graph algorithms and matrix mathematics are confirmed for relational operations. Finally, performance results illustrating the impact of these properties are presented and discussed.

## II. RELATIONS

The relational model, based on set theory, is a key mathematical foundation for SQL databases. The relational model effectively consists of relational algebra, relational calculus, and the structured query language (SQL) that balance the theoretical, implementation, and systems design aspects of databases. The relational model is well covered in the literature [Maier 1983, Codd 1990, Abiteboul 1995]; only the most relevant aspects of the model are reviewed here. Some of the more significant mathematical contributions of the relational model to databases include

(R1) Relations: a mathematical definition of database tables sufficient for their representation without constraining their implementation;
(R2) Query semantics: a mathematical definition of operations on relations sufficient for proving the correctness of database queries;
(R3) Proof of the equivalence of declarative and procedural syntaxes over the above definitions that has enabled the use of declarative semantics for database users and procedural semantics for database builders [Codd 1972].

Of these results, (R3) has been enormously important, but would not be possible without (R1) and (R2). (R3) has been critical to the success of SQL databases that follow the relational model. (R3) has enabled the successful coexistence of separate interfaces and languages for users and implementers, with the confidence that neither would create a fundamental mathematical contradiction for the other.

The relational model is based on balancing mathematical rigor with implementation practicality. Too much mathematical rigor burdens a database implementation with unnecessary mathematics. Too little mathematical rigor makes it is difficult to know if a database implementation will work. As with all good compromises, there have been advocates for improvement on both sides. As cited earlier, many new databases under the names of NoSQL and NewSQL differ from the relational model to meet new performance and analysis demands. Likewise, there is extensive mathematical work on modifications to the relational model to increase its mathematical rigor [Imieliński 1984a, Imieliński 1984b, Kanellakis 1989, Tsalenko 1992, Plotkin 1998, Priss 2006, van Emden 2006, Litak 2014, Hutchison 2016]. One motivation for increasing the mathematical rigor [Kelly 2012] is to align relations with well-established Zermelo-Fraenkel Choice (ZFC) set theory [Zermelo 1908, Fraenkel 1922] that is the foundation for a number of branches of mathematics.

The emerging diversity of databases has initiated a dialogue regarding the traditional relational model and the newer graph and matrix models. This dialogue is akin to the earlier declarative and procedural conversation that culminated in the relational model. This work seeks similar progress by demonstrating that an associative array model can provide

(A1) Associative arrays: a mathematical definition of database tables for SQL, NoSQL, and NewSQL databases that accurately describes their implementation;
(A2) Associative array algebra: a mathematical definition of database queries and computations that accurately describes the operations performed by SQL, NoSQL, and NewSQL databases;
(A3) Equivalence of relational and array syntaxes over the above definitions that enables the use of either in a SQL, NoSQL, or NewSQL database.

Of these results, (A3) has the most potential to impact polystore databases. Likewise, (A3) would not be possible without (A1) and (A2).

The mathematical challenge of creating an associative array model encompassing SQL, NoSQL, and NewSQL is reconciling their mathematical differences. SQL databases focus on set operations (subsets, unions, intersections), and the relational model is based on an elegant approach to set theory that provides only those attributes of formal set theory that are required for SQL databases. NoSQL and NewSQL databases focus on high performance data analysis (graph algorithms and matrix mathematics) that require mathematical properties such as associativity, commutativity, distributivity, identities, annihilators, and inverses. Reproducing the balance that led to the success of the relational model in another model is difficult. Detailed analysis of this balance leads down the same well-traveled path of those who have advocated for both more or less mathematical rigor in the relational model. Instead, just as Alexander solved the problem of the Gordian Knot, this paper asserts the desired outcome (relations are associative arrays) and the implications of this assertion are then addressed.

### III. GRAPHS AND MATRICES

The duality between graph algorithms and matrix mathematics (or sparse linear algebra) has been extensively covered in the literature and is summarized in the cited text [Kepner & Gilbert 2011]. This text has further spawned the development of the GraphBLAS math library standard (GraphBLAS.org)[Mattson 2013] that is described in the series of proceedings [Mattson 2014a, Mattson 2014b, Mattson 2015, Buluc 2015, Buluc 2016]. The essence of the graph algorithms and matrix mathematics duality are three operations: element-wise addition, element-wise multiplication, and matrix multiplication. In brief, an m×n matrix **A** is defined as a mapping from pairs of integers to values

$$\mathbf{A}: \{1,...,m\} \times \{1,...,n\} \rightarrow \mathbb{V}$$

where $\mathbb{V}$ is the set of values that form a semiring $(\mathbb{V},\oplus,\otimes,0,1)$[Kepner & Jansen 2016] with addition operation $\oplus$, multiplication operation $\otimes$, additive identity/multiplicative annihilator 0, and multiplicative identity 1. The construction of a sparse matrix is denoted

$$\mathbf{A} = \mathbb{A}(I,J,V)$$

where I, J, V are vectors of the rows, columns, and values of the nonzero elements of **A**.

Given m×n matrices **A**, **B**, and **C**, element-wise matrix addition (and its graph equivalent: weighted graph union) is denoted

$$\mathbf{C} = \mathbf{A} \oplus \mathbf{B}$$

or more specifically

$$\mathbf{C}(i,j) = \mathbf{A}(i,j) \oplus \mathbf{B}(i,j)$$

where $i \in \{1,...,m\}$ and $j \in \{1,...,n\}$. Element-wise matrix multiplication (and its graph equivalent, weighted graph intersection) is denoted

$$\mathbf{C} = \mathbf{A} \otimes \mathbf{B}$$

or more specifically

$$\mathbf{C}(i,j) = \mathbf{A}(i,j) \otimes \mathbf{B}(i,j)$$

For a m×l matrix **A**, l×n matrix **B**, and m×n matrix **C**, matrix multiplication (and its graph equivalent, multisource weighted breadth-first search) combines addition and multiplication and is written

$$\mathbf{C} = \mathbf{A} \oplus.\otimes \mathbf{B} = \mathbf{A}\ \mathbf{B}$$

or more specifically

$$\mathbf{C}(i,j) = \oplus_k\ \mathbf{A}(i,k) \otimes \mathbf{B}(k,j)$$

where $k \in \{1,...,l\}$. Finally, the matrix transpose (and its graph equivalent, graph edge reversal) is denoted

$$\mathbf{A}(j,i) = \mathbf{A}^\mathsf{T}(i,j)$$

The above operations have been found to enable a wide range of graph algorithms and matrix mathematics while also preserving the required vector-space properties [Heaviside 1887, Peano 1888] such as commutativity

$$\mathbf{A} \oplus \mathbf{B} = \mathbf{B} \oplus \mathbf{A}$$
$$\mathbf{A} \otimes \mathbf{B} = \mathbf{B} \otimes \mathbf{A}$$
$$(\mathbf{A}\ \mathbf{B})^\mathsf{T} = \mathbf{B}^\mathsf{T}\ \mathbf{A}^\mathsf{T}$$

associativity

$$(\mathbf{A} \oplus \mathbf{B}) \oplus \mathbf{C} = \mathbf{A} \oplus (\mathbf{B} \oplus \mathbf{C})$$
$$(\mathbf{A} \otimes \mathbf{B}) \otimes \mathbf{C} = \mathbf{A} \otimes (\mathbf{B} \otimes \mathbf{C})$$
$$(\mathbf{A}\ \mathbf{B})\ \mathbf{C} = \mathbf{A}\ (\mathbf{B}\ \mathbf{C})$$

distributivity

$$\mathbf{A} \otimes (\mathbf{B} \oplus \mathbf{C}) = (\mathbf{A} \otimes \mathbf{B}) \oplus (\mathbf{A} \otimes \mathbf{C})$$
$$\mathbf{A}\ (\mathbf{B} \oplus \mathbf{C}) = (\mathbf{A}\ \mathbf{B}) \oplus (\mathbf{A}\ \mathbf{C})$$

and the additive and multiplicative identities

$$\mathbf{A} \oplus \mathbb{0} = \mathbf{A}$$
$$\mathbf{A} \otimes \mathbb{1} = \mathbf{A}$$
$$\mathbf{A}\ \mathbb{I} = \mathbf{A}$$

where $\mathbb{0}$ is a matrix of all 0, $\mathbb{1}$ is a matrix of all 1, and $\mathbb{I}$ is a matrix with 1 along its diagonal. Furthermore, these matrices possess a multiplicative annihilator

$$\mathbf{A} \otimes \mathbb{0} = \mathbb{0}$$
$$\mathbf{A}\ \mathbb{0} = \mathbb{0}$$

Their corresponding inverses may also exist

$$\mathbf{A} \oplus -\mathbf{A} = \mathbb{0}$$
$$\mathbf{A}(i,j) \otimes \mathbf{A}(i,j)^{-1} = \mathbb{1}$$
$$\mathbf{A}\ \mathbf{A}^{-1} = \mathbb{I}$$

when $(\mathbb{V},\oplus,0)$ and $(\mathbb{V},\otimes,1)$ are groups (i.e., have inverses) [Galois 1832].

Most significantly, the properties of matrices are determined by the properties of the set of values $\mathbb{V}$. In other words, the properties of $\mathbb{V}$ determine the properties of the corresponding matrices. The above properties are required for the development and implementation of data analysis algorithms.

## IV. ASSOCIATIVE ARRAYS

Associative arrays can be rigorously built up from ZFC set theory, groups, and semirings, culminating with the observation that linear algebra is a specialization of associative array algebra. How associative arrays encompass graphs, matrices, NoSQL, and NewSQL is described extensively in [Kepner & Jansen 2016] and is only summarized here.

As described earlier, sparse matrices are a common representation used for both graphs and linear algebra. The standard definition of sparse matrices requires generalization to encompass the tables found in SQL, NoSQL, and NewSQL databases. The primary difference between a matrix and an associative array is the specification of the row and column indices. In a matrix, the row and column indices are drawn from the sets of integers $\{1,...,m\}$ and $\{1,...,n\}$. Associative array row and column "keys" can be drawn from any strict, totally ordered set (i.e., any uniquely sortable set). Associative array row and column keys can be negative numbers, real numbers, or character strings. The true dimensions of an associative array are often very large (e.g., all possible finite strings). Instead, the size of an associative array is more commonly used and is defined as the number of nonzero rows, $\underline{m}$, and the number of nonzero columns, $\underline{n}$. An equally important quantity is the number of nonzeros in an associative array, which is denoted by the function nnz(). The size and nnz of an associative array can change through the course of a calculation. There are no size constraints on associative array operations. Element-wise addition, element-wise multiplication, and array multiplication are valid for combinations of associative arrays of any size.

Associative arrays derive much of their power from their ability to represent data intuitively in easily understandable tables. Consider the list of songs and the various features of those songs shown in Figure 5. The tabular arrangement of the data shown in Figure 5 is an associative array (denoted **A**). This arrangement is similar to those widely used in spreadsheets and databases. Figure 5 illustrates two properties of associative arrays that may differ from other two-dimensional arrangements of data. First, each row key and each column key in **A** unique, to allow rows and columns to be queried efficiently. Second, associative arrays do not store rows or columns that are entirely empty, to allow insertion, selection, and deletion of data to be performed by associative array addition, multiplication, and products. These properties are what makes **A** an associative array and allows **A** to be manipulated as a spreadsheet, database, matrix, or graph.

|              | Artist   | Date       | Duration | Genre      |
|--------------|----------|------------|----------|------------|
| 053013ktnA1  | Bandayde | 2013-05-30 | 5:14     | Electronic |
| 053013ktnA2  | Kastle   | 2013-05-30 | 3:07     | Electronic |
| 063012ktnA1  | Kitten   | 2010-06-30 | 4:38     | Rock       |
| 082812ktnA1  | Kitten   | 2012-08-28 | 3:25     | Pop        |

Figure 5. Tabular arrangement of a list of songs and the various features of those songs into an associative array **A**. The array **A** is an associative array because each row label (or row key) and each column label (or column key) in **A** is unique. The size of the associative array is $\underline{m}$ = 4 and $\underline{n}$ = 4.

## V. RELATIONS AS ASSOCIATIVE ARRAYS

A first step in adapting the relational model to associative arrays is to define a relation in terms of associative arrays. This step is done by asserting a relation *is* an associative array and considering the implications of the assertion. Operationally, asserting that relations are associative arrays means that the row keys of an associative array are arbitrary but distinct at the time of input and output of a relational operation. Using this definition, some of the implications can be illustrated by a series of common questions about relations, specifically, whether or not relations are sets, tuples, indices, ordered, multisets (bags), or sequences.

Are relations ZFC sets? Relations in the traditional relational model require some, but not all, properties of ZFC set semantics. The values of associative arrays are ZFC sets. The keys of associative arrays are ZFC sets. Expressing relations as associative arrays means that they adhere to ZFC set semantics.

Are relations tuples? A row of an associative array is mathematically a row vector. Mathematically, tuples are also vectors so relations are tuples.

Are relations indices? In the past, it has been efficient in both space and time if a relation can be represented as a tuple of integer indices that connect to values in a table. Today, this implementation guidance is less important and it is mathematically more flexible to treat relations as tuples of their actual values, which is how they are defined in associative arrays.

Are relations ordered sets? Mathematically, ordering of rows or columns is not required for either relations or associative arrays. However, as a practical matter, ordering is required for real database tables, and there is no negative mathematical consequence for requiring rows and columns to be ordered sets. Thus, associative array rows and columns are ordered sets.

Are relations multisets (bags)? Identical rows are a reality in many databases, implying that relations are multi-sets. The row key of an associative array distinguishes rows with identical values.

Are relations sequences? A practical approach to implementing multiple identical rows is to view relations as a mathematical sequence instead of a set. In a sequence, each row is paired with a number that sets the order of the rows; hence, the term sequence ID in SQL databases. A sequence ID is effectively equivalent to the row key in an associative array. NoSQL databases embrace this view to the point of fully exposing the unique sequence ID as a user-controlled parameter.

Defining relations as associative arrays provides new answers to the above questions. However, new questions arise that also must be addressed. Primary amongst these are the differences among 0, $\varnothing$, null, empty entries, and empty rows and columns. To provide the necessary mathematical properties for matrix calculations, associative arrays follow the conventions set by sparse matrices that define 0 as the non-stored element. More specifically, the value corresponding to the ⊕ identity and the ⊗ annihilator is the non-stored element. Because of its mathematical properties, the 0 element is unique and there is no distinction between 0 and "no data" or null. As a practical matter, when it is desired to distinguish between these states, usually a workaround can be found. A unique 0 is useful as it does not require that exceptions be defined for every operation.

## VI. Queries as Associative Array Algebra

Relational algebra and SQL have defined a wide range of operations that are useful for executing queries on relations. Some of these operations are union ∪, intersection ∩, set difference \, Cartesian product ×, project Π, rename ρ, select σ, natural join ⋈, equijoin ⋈$_k$, theta join ⋈$_θ$, left outer join ⟕, right outer join ⟖, full outer join ⟗, antijoin ▷, extended projection, and aggregation. In discussions of the relational model, the list of operations most commonly discussed include union ∪, intersection ∩, set difference \, project Π, rename ρ, select σ, and theta join ⋈$_θ$.

In practice, all computations are restricted to the nonzero rows and nonzero columns of the associative array representation of relations. . Likewise, since the row keys in an associative array representation of a relation are arbitrary, it is assumed that wherever convenient the row keys of any associative array can be made distinct. Thus, it is common for there to be no operations that require the comparison of two arbitrary values. In many computations, the only operations that need to be specified are the identities

$$v \oplus 0 = v \qquad v \otimes 1 = v$$

and the additive inverse and multiplicative annihilator

$$v \oplus -v = 0 \qquad v \otimes 0 = 0$$

where $v \in \mathbb{V}$. Results that can be proven under the above conditions will hold for a wide variety of ⊕ and ⊗ operations.

### A. Equivalence

In dealing with any new data representation, the first step is to define when two representations are equivalent [Howe 2005]. Relational equivalence for associative arrays is denoted

$$\mathbf{A} \sim \mathbf{B}$$

and implies every row in **A** has an identical row in **B**, and every row in **B** has an identical row in **A**. This definition allows multiple identical rows. A stronger version further requires exactly the same number of identical rows in **A** and **B**. Equivalence can be computed via the equivalency permutation array **P** of the nonzero rows in **A** to the nonzero rows in **B** where $\mathbf{P}(i_A, i_B) = 1$ (and 0 otherwise) if row $\mathbf{A}(i_A,:)$ is the same as row $\mathbf{B}(i_B,:)$. **P** can be computed by using a variety of notational conventions

$$\mathbf{P} = \mathbb{I}_A \ (\mathbf{A} \oplus .\otimes \mathbf{B}^T) \ \mathbb{I}_B$$
$$= \mathbb{I}_A \ (\mathbf{A} \ \&.= \mathbf{B}^T) \ \mathbb{I}_B$$

where ⊕ is &, ⊗ is =, and

$I_A$ is the set of nonzero rows in **A**
$\mathbb{I}_A = \mathbb{A}(I_A, I_A, 1)$ is the identity array over $I_A$
$I_B$ is the set of nonzero rows in **B**
$\mathbb{I}_B = \mathbb{A}(I_B, I_B, 1)$ is the identity array over $I_B$

or more specifically

$$\mathbf{P}(i_A, i_B) = \&_j \ (\mathbf{A}(i_A, j) = \mathbf{B}(i_B, j))$$

where $i_A \in I_A$ and $i_B \in I_B$. Likewise, **P** can computed as

$$\mathbf{P}(i_A, i_B) = \delta(\mathbf{A}(i_A,:), \mathbf{B}(i_B,:))$$

where δ(,) is the Kronecker delta function. If every nonzero row in **A** has a nonzero row in **P** and if every row in **B** has a nonzero column in **P**, then

$$\mathbf{A} \sim \mathbf{B}$$

Using the convention of restricting to the nonzero rows of **A** and **B**, **P** can also be computed simply as

$$\mathbf{P} = \mathbf{A}\mathbf{B}^T$$

where ⊕.⊗ is &.= or δ(,) is implied. Likewise, by the transpose identity

$$\mathbf{P}^T = \mathbf{B}\mathbf{A}^T$$

The stronger version of equivalence can obtained by imposing the further requirement that if **P** is stripped of its row and column keys, it forms a symmetric matrix where

$$\mathbf{P}^T = \mathbf{P}$$

Using this definition of equivalence allows most relational operations to be defined in terms of variations on the construction of the permutation matrix **P**.

### B. Project

The project operation picks sets of J columns from a relation **A** and is typically written in relational algebra as

$$\Pi_J(\mathbf{A})$$

The SQL equivalent is

SELECT J(1),...,J(n) FROM **A**

or simply

$$\mathbf{A}._{J(1),...,J(n)}$$

In terms of associative array algebra, project can be accomplished via many expressions, such as

$$\mathbf{A} \oplus .\otimes \mathbb{A}(J,J,1) \quad \text{or} \quad \mathbf{A} \ \mathbb{I}(J,J) \quad \text{or} \quad \mathbf{A} \ \mathbb{I}(J) \quad \text{or} \quad \mathbf{A}(:,J)$$

given the shorthand notation for the identity array

$$\mathbb{A}(J,J,1) = \mathbb{I}(J,J) = \mathbb{I}(J)$$

and the Matlab notation $\mathbf{A}(:,J)$ for column selection.

### C. Rename

The rename operation picks columns J from a relation **A** and assigns them new names J'. Rename is written in relational algebra as

$$\rho_{J/J'}(\mathbf{A})$$

The SQL equivalent is

SELECT J(1),...,J(n) AS J'(1),...,J'(n) FROM **A**

In associative array algebra, rename can be accomplished with the many expressions, such as

$$\mathbf{A} \oplus .\otimes \mathbb{A}(J,J',1) \quad \text{or} \quad \mathbf{A} \ \mathbb{I}(J,J')$$

### D. Union

The union operation selects all the distinct rows in two relations **A** and **B** and is written in relational algebra as

$$\mathbf{A} \cup \mathbf{B}$$

The SQL equivalent is

SELECT ∗ FROM **A** UNION SELECT ∗ FROM **B**

In associative array algebra, using the convention of distinct row keys for nonzero rows, union can be written as

$$\mathbf{A} \oplus \mathbf{B}$$

*E. Intersection*

The intersection operation combines the common rows in two relations **A** and **B** and is written in relational algebra as

$$\mathbf{A} \cap \mathbf{B}$$

The SQL equivalent is

SELECT ∗ FROM **A** INTERSECT SELECT ∗ FROM **B**

In associative array algebra, using the equivalence permutation array, intersection can be computed with the following expressions

$$\mathbf{PB} \quad \text{or} \quad \mathbf{P}^T\mathbf{A}$$

*F. Set Difference*

Set difference returns the rows in relation **A** that are not found in relation **B** and is written in relational algebra as

$$\mathbf{A} \setminus \mathbf{B}$$

The SQL equivalent is

SELECT ∗ FROM **A** EXCEPT SELECT ∗ FROM **B**

In associative array algebra, assuming the additive inverse v ⊕ -v = 0, intersection can be written using the equivalence permutation array as

$$\mathbf{A} \oplus \mathbf{-PB}$$

*G. Select*

The select operation returns all rows in the relation **A** that satisfy a function φ on the subset of columns J

$$\sigma_{\varphi(J)}(\mathbf{A})$$

The SQL equivalent is

SELECT ∗ FROM **A** WHERE φ(**A**.$_{J(1),...,J(n)}$)

In associative array algebra, select can be written using the select permutation array

$$\mathbf{PA}$$

where

$$\mathbf{P} = (\varphi(\mathbf{A}(:,J)) \; \varphi(\mathbf{A}(:,J))^T) \; \otimes \; \mathbb{I}_\mathbf{A}$$

or

$$\mathbf{P} = \mathbb{I}(\varphi(\mathbf{A}(:,J)))$$

The function φ can be any function on the J columns of a row of an associative array that produces either a 0 or a 1 (i.e., ⊕.θ ∈ {0,1}).

*H. Theta Join*

A theta join returns the rows of two relations **A** and **B** joined where they satisfy the function θ on the J columns of **A** and J' columns of **B** and is written in relational algebra as

$$\mathbf{A} \bowtie_{\theta(J,J')} \mathbf{B}$$

The SQL equivalent is

SELECT ∗ FROM **A**, **B** WHERE θ(**A**.$_{J(1),...,J(n)}$,**B**.$_{J'(1),...,J'(n)}$)

In associative array algebra, select can be written using the θ permutation array as

$$\mathbf{PB} \oplus \mathbf{PP}^T\mathbf{A} \quad \text{or} \quad \mathbf{P}^T\mathbf{A} \oplus \mathbf{P}^T\mathbf{PB}$$

where

$$\mathbf{P} = \mathbb{I}_\mathbf{A} \; (\mathbf{A}(:,J) \oplus.\otimes \mathbf{B}(:,J')^T) \; \mathbb{I}_\mathbf{B}$$
$$= \mathbb{I}_\mathbf{A} \; (\mathbf{A}(:,J) \; \theta \; \mathbf{B}(:,J')^T) \; \mathbb{I}_\mathbf{B}$$

The function θ can be any function on two rows of an associative array that produces either a 0 or a 1 (i.e., θ ∈ {0,1}). If the operation is restricted to the nonzero rows of **A** and **B**, then the $\mathbb{I}_\mathbf{A}$ and $\mathbb{I}_\mathbf{B}$ terms can be dropped and the θ permutation array can be written as

$$\mathbf{P} = \mathbf{A}(:,J) \; \theta \; \mathbf{B}(:,J')^T$$

*I. Extended Projection*

An extended projection applies a function φ on the subset of columns J of a relation **A** and returns the output of that function as a new relation with a column key j'. Extended projection is written in relational algebra as

$$_{j'}\Pi_{\varphi(J)}(\mathbf{A})$$

The SQL equivalent is

SELECT φ(**A**.$_{J(1),...,J(n)}$) AS j' FROM **A**

In associative array algebra, extended projection can be written as

$$\mathbf{A} \; \oplus.\otimes \; \mathbb{I}(J,j')$$

where ⊕.⊗ = φ. The function φ can be any function on a row of an associative array.

*J. Aggregation*

The aggregation operations applies an aggregate function *f* on all the values of column j' of relation **A** that share a common value in column j. Aggregation is written in relational algebra as

$$_{j'}\mathrm{G}_{f(j)}(\mathbf{A})$$

The SQL equivalent is

SELECT $f_{j'}$ FROM **A** GROUP BY j

In associative array algebra, aggregation can be written as

$$\mathbf{P} \; f.\otimes \; \mathbf{A}(:,j')$$

where P is the permutation array defined by cross-correlating column j with itself

$$\mathbf{P} = \mathbb{I}_\mathbf{A} \; (\mathbf{A}(:,j) \oplus.\otimes \mathbf{A}(:,j)^T) \; \mathbb{I}_\mathbf{A}$$
$$= \mathbb{I}_\mathbf{A} \; (\mathbf{A}(:,j) \oplus.= \mathbf{A}(:,j)^T) \; \mathbb{I}_\mathbf{A}$$

The function *f* can be any binary commutative function on a column of an associative array.

VII. PROPERTIES AND PERFORMANCE

Having expressed the main relational operations in terms of associative array algebra, the mathematical properties necessary for graph and matrix computation can be checked. The results of this analysis are summarized in Tables 1 and 2. The derivation of all of these properties is beyond the scope of this paper, but as an example, perhaps the most important property, distributivity is derived in the context of relational renaming over union and intersection. These operations are the closest direct analogs to array multiplication and element-wise addition.

Showing that renaming distributes over union is computed as follows

$$\rho_{J/J'}(\mathbf{A} \cup \mathbf{B}) \sim (\mathbf{A} \oplus \mathbf{B})\, \mathbb{I}(J,J')$$
$$= \mathbf{A}\, \mathbb{I}(J,J') \oplus \mathbf{B}\, \mathbb{I}(J,J')$$
$$\sim \rho_{J/J'}(\mathbf{A}) \cup \rho_{J/J'}(\mathbf{B})$$

Showing that renaming distributes over intersection is computed as follows

$$\rho_{J/J'}(\mathbf{A}) \cap \rho_{J/J'}(\mathbf{B}) \sim ((\mathbf{A}\, \mathbb{I}(J,J'))\, (\mathbf{B}\, \mathbb{I}(J,J'))^T)\, \mathbf{B}\, \mathbb{I}(J,J')$$
$$= ((\mathbf{A}\, \mathbb{I}(J,J'))\, (\mathbb{I}(J,J')^T\, \mathbf{B}^T))\, \mathbf{B}\, \mathbb{I}(J,J')$$
$$= ((\mathbf{A}\, (\mathbb{I}(J,J')\, \mathbb{I}(J',J)))\, \mathbf{B}^T)\, \mathbf{B}\, \mathbb{I}(J,J')$$
$$= (\mathbf{A}\, \mathbb{I}(J)\, \mathbf{B}^T)\, \mathbf{B}\, \mathbb{I}(J,J')$$
$$= (\mathbf{A}\mathbf{B}^T)\, \mathbf{B}\, \mathbb{I}(J,J')$$
$$= (\mathbf{P}\mathbf{B})\, \mathbb{I}(J,J')$$
$$\sim \rho_{J/J'}(\mathbf{A} \cap \mathbf{B})$$

Because the above derivations use the corresponding properties of associative arrays, the results can be more general than the relational algebra would suggest. Specifically, distributivity would still hold if rename modified the values in a manner consistent with associative array multiplication. Likewise, distributivity would still hold if union and intersection modified the values in a manner consistent with element-wise addition.

Table 1. Identity, annihilator, and inverse properties of relational operations in terms of associative arrays. Unary functions with parameters are treated as binary functions. The potential performance impact of the elimination of an operation via the recognition of one of these properties is typically $O(\mathrm{nnz}(\mathbf{A}))$.

| Operation | Identity | Annihilator | Inverse |
|---|---|---|---|
| Project | $\mathbb{I}$ | $\mathbb{0}$ | |
| Rename | $\mathbb{I}$ | $\mathbb{0}$ | |
| Union | $\mathbb{0}$ | | |
| Intersection | | $\mathbb{0}$ | |
| Difference | $\mathbb{0}$ | | $\mathbf{A}$ |
| Select | $\varphi = 1$ | $\varphi = 0$ | |
| Theta Join | | $\mathbb{0}$ | $\mathbf{A}$ |
| Extended Projection | $J = j'$, $\varphi(1,v) = v$ | $\varphi = 0$ | $\varphi = 1$ |
| Aggregation | $\mathbf{A}(:,j)$ is unique, $f(0,v) = v$ | $f(0,v) = 0$ | $f(0,v) = 1$ |

Table 2. Commutativity, associativity, and distributivity properties of relational operations on associative arrays. Unary functions with parameters are treated as binary functions. The potential performance impact of elimination of an operation via the recognition of one of these properties is typically $O(\mathrm{nnz}(\mathbf{A}))$.

| Operation | Commutativity | Associativity | Distributivity |
|---|---|---|---|
| Projection | no | yes | $\cup, \cap, \setminus$ |
| Rename | no | yes | $\cup, \cap, \setminus$ |
| Union | yes | yes | $\cap$ |
| Intersection | yes | yes | $\cup$ |
| Difference | no | no | |
| Select[1] | | | |
| Theta Join[2] | yes | | |

[1] Assumes corresponding property in select function $\varphi$.

[2] Assumes corresponding property in join function $\theta$.

One of the benefits of the properties in Table 1 is the ability to eliminate operations if the appropriate identity can be recognized. Likewise, the properties in Table 2 allow operations to be reordered to reduce execution time. These properties are particularly useful in the polystore context when selecting the optimal database to perform an operation. Figures 6 and 7 show the relative execution time impact of exploiting associativity and distributivity as a function of the size of the associative arrays. These experimental measurements were conducted using the D4M (d4m.mit.edu) implementation of associative arrays. A fixed 4096×4096 associative array $\mathbf{A}$ was multiplied with square associative arrays $\mathbf{B}$ and $\mathbf{C}$ that varied in size. All of the associative arrays were randomly generated with an average of 8 nonzero entries per row or column, which is consistent with many graph applications. The results show the potential performance benefits of exploiting the distributive and associative properties. Thus, the kinds of query optimizations that are found in many databases systems can be applied to a broad set of computations. These optimizations are important for polystores as they allow the movement of computations and data to the appropriate databases.

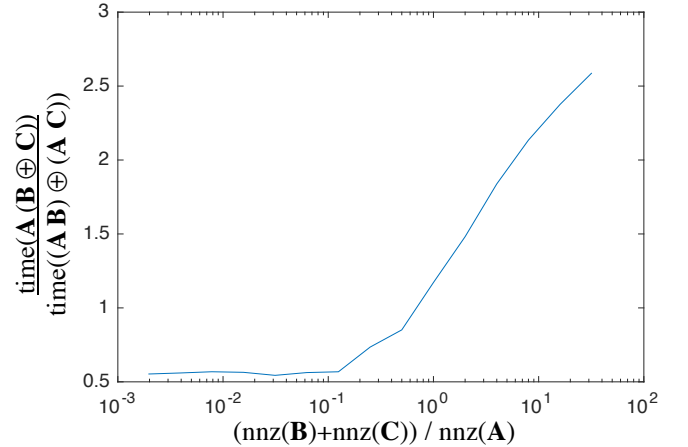

Figure 6. Relative execution time of $\mathbf{A}\,(\mathbf{B} \oplus \mathbf{C})$ vs $(\mathbf{A}\,\mathbf{B}) \oplus (\mathbf{A}\,\mathbf{C})$ as a function of the size of $\mathbf{B}$ and $\mathbf{C}$ as compared to $\mathbf{A}$. $(\mathbf{A}\,\mathbf{B}) \oplus (\mathbf{A}\,\mathbf{C})$ is much faster than $\mathbf{A}\,(\mathbf{B} \oplus \mathbf{C})$ when $\mathbf{A}$ is smaller than $\mathbf{B}$ and $\mathbf{C}$.

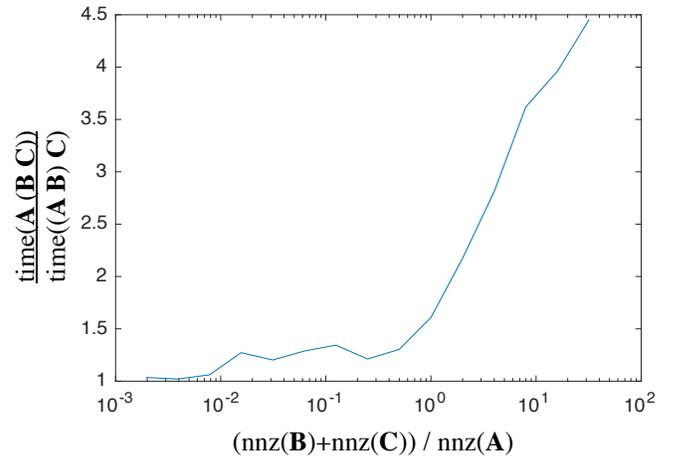

Figure 7. Relative execution time of **A** (**B C**) vs (**A B**) **C** as a function of the size of **B** and **C** as compared to **A**. (**A B**) **C** is much faster than **A** (**B C**) when **A** is smaller than **B** and **C**.

VIII. SUMMARY AND FUTURE WORK

The success of SQL, NoSQL, and NewSQL databases is a reflection of their ability to provide significant functionality and performance benefits for specific domains: transactions, internet search, and data analysis. The BigDAWG polystore seeks to provide a mechanism to allow applications to transparently achieve the benefits of diverse databases while insulating applications from the details of these diverse databases. Associative arrays provide a common approach to the mathematics found in different databases: sets (SQL), graphs (NoSQL), and matrices (NewSQL). This work presents the SQL relational model in terms of associative arrays and identifies the key mathematical properties of NoSQL and NewSQL that are preserved within SQL. These properties include associativity, commutativity, distributivity, identities, annihilators, and inverses. Performance measurements on distributivity and associativity show the impact these properties can have on associative array operations. These results demonstrate that associative arrays can provide a model for polystores to leverage mathematical properties across databases to optimize the exchange of data and queries.

Future work in this area will focus on a complete set of proofs for the aforementioned relational operations, detailed analysis of optimizations, and the potential application of uncertainty quantification to database queries.

ACKNOWLEDGMENTS

The authors wish to acknowledge the following individuals for their contributions: Michael Stonebraker, Sam Madden, Bill Howe, David Maier, Chris Hill, Alan Edelman, Charles Leiserson, Dave Martinez, Sterling Foster, Paul Burkhardt, Victor Roytburd, Bill Arcand, Bill Bergeron, David Bestor, Chansup Byun, Mike Houle, Matt Hubbell, Mike Jones, Anna Klein, Pete Michaleas, Lauren Milechin, Julie Mullen, Andy Prout, Tony Rosa, Sid Samsi, and Chuck Yee.